\documentclass{ws-procs9x6}

\begin{document}

\title{Tachyon Dynamics and Brane Cosmology}

\author{GARY SHIU}

\address{Department of Physics, University of Wisconsin, Madison, WI 53706, 
USA \\
E-mail: shiu@physics.wisc.edu}


\maketitle

\abstracts{We discuss the cosmological implications of
some recent advances in understanding 
the dynamics of tachyon condensation in string theory.}

The brane world scenario\cite{HW,ADD,ST,Lykken,Ovrut,RS}
radically alters our traditional view of how
the Standard Model interacts with gravity, and so it
provides a new framework to think about
problems in cosmology. 
In this regard, inflation is a particularly
promising arena: although
its predictions of the flatness of the universe and an almost
scale invariant density perturbation spectrum are in good
agreement with observations, it is still a paradigm
in search of a model.

The superluminal expansion of inflation is usually pictured as
driven by the potential
of a rolling scalar field, i.e., the inflaton.
In the brane world scenario, a natural candidate for the inflaton
      is the brane mode whose expectation value describes
the inter-brane separation\cite{DvaliTye}.
 The dynamics of inflation is therefore governed by the interactions
between D-branes, which is mediated by the exchanges of closed strings.
This idea of {\it brane inflation}\cite{DvaliTye} has been
applied to construct
inflationary models arising from the collision of branes and 
antibranes\cite{DvaliShafi,Burgess}, as well
as branes intersecting at 
angles\cite{Herdeiro,Kyae,Garcia-Bellido,Blumenhagen:Inflation,Dasgupta,Gomez-Reino}.
The potential which drives inflation
(at least during the slow-roll epoch when the branes are far apart)
is well studied in string theory\cite{Polchinski}. 
When the branes are sufficiently close to one another, tachyons 
develop\cite{BanksSusskind}
and the universe enters into a hybrid inflation\cite{hybrid} regime. The
slow change in inter-brane separation is followed by a 
quick roll down the tachyonic direction.
Eventually, inflation ends when the brane collide, heating the universe that
starts the big-bang.

What are the testable signatures of this scenario?
First of all, in addition to the static potential,
there is generically a velocity dependent potential between 
D-branes (see, e.g.,\cite{Polchinski}).
Since the branes are moving
during inflation, the velocity-dependent potential
may 
leave an imprint on cosmological measurements\cite{braneinf}.
More concretely, this new contribution
can 
modify the ratio of the tensor to scalar fluctuations,  
and so a powerful test of such effect from brane inflation
is to observe the violation of
the
inflationary consistency condition\footnote{For a review on
reconstructing the inflation potential and the consistency
condition, see, e.g.,\cite{reconstruct} and references therein.}
.

The tachyon which appears towards the end of brane inflation has even more
interesting cosmological consequences.
In particular, 
there has been some recent 
progress in constructing the supersymmetric Standard Model
from intersecting 
branes\cite{CSU1,CSU2,CSU3}.
In these models, the brane configurations are rather special. A natural
question is: why (and how) are these special vacua
selected? In a cosmological setting,
it is likely that the initial configuration of branes are not perfectly
stable, e.g.,
the branes can intersect at non-supersymmetric 
angles\cite{Berlin1,angelantonj,Madrid1,Madrid2,Berlin2}, or
there
could perhaps be additional pairs of branes and anti-branes in the early
universe.
The tachyon instability drives the time evolution
of the unstable system to a nearby stable configuration, 
so we can think of the supersymmetric models in\cite{CSU1,CSU2}
as the endpoints
of some rolling tachyons. 
It is therefore both interesting and relevant to explore the 
implications and prospects of tachyons in 
cosmology\cite{Tachyon1,Tachyon2,Frolov,Kofman}.

The recent advances in tachyon condensation in string
theory\cite{Sen:review}
provide several of the ingredients needed to analyze its
cosmological implications.
According to the conjecture of Sen\cite{Sen:review}, 
when the tachyon which appears in the
worldvolume of a non-BPS brane or a brane-antibrane pair condense, the
endpoint of the condensation is a closed string vacuum without
any open string excitations.
More recently, time-dependent solutions of 
the rolling tachyon have been
obtained in\cite{sen1,sen2,sen3}. Related solutions have also been
discussed in\cite{Sugimoto,Minahan} from considerations of the 
boundary string field theory\cite{BSFT} 
actions for non-BPS branes\cite{Kutasov}
and brane-antibrane pairs\cite{Kraus,Takayanagi}.
These works suggested that 
the endpoint of tachyon condensation
can be described by a pressureless gas with non-zero energy density, known as
tachyon matter\footnote{This pressureless tachyon matter is related to
the S-brane solutions in\cite{s-branes}.}.
Motivated by the similarity of the tachyon matter with
a non-relativistic dust\cite{sen1,sen2,sen3,Gibbons},
its cosmological constraints (in particular, 
the question of whether it
is a viable candidate for the cosmological dark matter) were
examined in\cite{Tachyon1,Frolov}.
Unlike quintessence,
tachyon matter can cluster on very small
scales 
and thus can play the role of dark matter.
However, it is easy to estimate
that the tachyon matter density is many orders
of magnitude too big to be compatible with present day cosmological
observations\cite{Tachyon1}. This poses a severe overabundance
problem, especially for brane inflationary models, because there is not
enough number of e-foldings for the
tachyon arising from the last collision of branes to be
inflated away.

Some possible solutions to this overabundance problem were explored 
in\cite{Tachyon2}.
In particular, the overabundance
problem can be solved if almost all of the 
tachyon energy
is drained to heating the universe (via coupling to the inflaton
and matter fields) and gravitational radiation, 
while the rest of the enegy goes
to the formation of a cosmic string network 
at the
end of inflation.
As discussed in\cite{Strominger:reheat,sen:divergent,Okuda,Chen}, if 
quantum effects are taken into account, we expect 
that the tachyon
arising from the annihilation of brane-antibrane will decay rapidly to closed 
strings.
However, in more realistic models, there are branes that survive the
annihilation, i.e., the branes containing the Standard Model.
The tachyon also couples to the open strings on the Standard Model branes, so
some fraction of the energy will be converted to normal radiation as well. 
This is promising because
it was shown in\cite{Tachyon2} that as long 
as the decay width of the tachyon
is not decreasing sufficiently fast with time, the tachyon matter density 
becomes insignificant rather rapidly.

Interestingly, the cosmic string network produced at the end of inflation
offers an exciting opportunity of probing brane world physics from
cosmological observations.
The vacuum manifold of the tachyon field $T$ is non-trivial, and in
fact supports stable defects (lower dimensional BPS branes)
of even codimensions\cite{Ktheory}.
Cosmologically, we expect these defects are produced\cite{defects}
because the tachyon field $T$ can take
different values at different spatial points. The existence of a
particle horizon implies that $T$ cannot be correlated on scales
larger than the horizon length $H^{-1}$ where $H$ is the Hubble
parameter during inflation. Therefore, defects will generically be
produced via the Kibble mechanism
with a density of order one per Hubble volume.
However, the sizes of the compact dimensions are much
smaller than the Hubble scale and so the defects
that are formed cosmologically will have their codimensions
along the non-compact dimensions. Since the stable defects
have even codimensions, only cosmic strings will be formed via
the Kibble mechanism whereas the cosmological production of domain walls and
monopole-like objects are heavily suppressed.
The Kibble mechanism implies an initial 
density $\sim H^2 M_s^2 \sim M_s^6/M_P^2$ in cosmic strings at the end of 
brane inflation.
Intercommutation of intersecting cosmic strings and the decay of string loops
(to gravitational radiation) causes the density of the cosmic string 
network (CSN) to
approach the scaling solution\cite{CSN} $\Omega_{CSN} \sim G \mu \sim M_s^2/M_P^2$, 
which would be acceptable.
Since the cosmic strings produced in this scenario\cite{Sarangi} have 
$10^{-6} \geq G \mu \geq 10^{-10}$, this scenario is provisionally
not ruled out\cite{Jones,Sarangi} by the present data\footnote{As discussed 
in \cite{Bouchet}, the present CMB data
can easily accomodate up to 20\% of the CMB anisotropy.}.
However, future data from 
MAP and PLANCK together with measurements from gravitational wave detectors
and perhaps pulsar timing will allow us to test this idea. 
The size of these cosmic string effects depends on the 
string value $M_s$. Therefore, cosmological measurements in the near future
may well provide a powerful way
to experimentally determine the value of $M_s$.

To conclude, inflation provides an exciting prospect of
testing the brane world idea from cosmological measurements.
Moreover, the dynamics of tachyon condensation could play an
important role in cosmology, especially in inflationary models
whose origin is the interaction between branes.
The recent developments in tachyon condensation in
string theory have allowed us to address some of its cosmological
implications, 
but many interesting problems remain to be explored, e.g.,

\noindent $\bullet$
As the tachyon field does not oscillate around the minimum of the 
potential, it is important to reexamine the mechanism of
reheating in this context\cite{Tachyon2,Kofman,Cline}.
It would also be interesting to carry out a direct string theoretical 
calculation of reheating by computing the couplings of the tachyon
to the Standard Model particles as well as closed strings.
This would allow us to better estimate 
what fraction of the tachyon energy goes to 
normal radiation, gravitational radiation, tachyon matter, and cosmic
strings, respectively.

\noindent $\bullet$ Using the p-adic string as a toy model,
time-dependent solutions of a rolling tachyon whose action involves an 
infinite number of
derivative terms have recently been studied in\cite{Moeller}.
It is worthwhile to explore the cosmology
of this model since this would allow us to address issues
that require an understanding of higher derivative terms in the tachyon
action, e.g.,
the problem of caustics formation
as recently pointed out in\cite{caustic}.

\noindent $\bullet$ Another interesting direction is to
investigate the properties of the
tachyon matter arising from D-branes intersecting at non-supersymmetric
angles.
Besides the 
relevance to the brane world models 
in\cite{CSU1,CSU2,CSU3,Berlin1,angelantonj,Madrid1,Madrid2,Berlin2}, 
such tachyon matter
may also 
exhibit new qualitative features.
Unlike the tachyon
on a non-BPS brane or a brane-antibrane pair, the potential
for the tachyon from intersecting
D-branes is not universal but depends
on the specific conformal field theory of the open strings.

\noindent $\bullet$ It would be interesting to examine the role that the 
rolling tachyons 
could play in
the dynamics of brane gas\cite{Alexander},
and also the new features that arise when
we embed the brane inflationary scenario in
various warped compactifications\cite{Ovrut,RS,warp,Giddings,Verlinde}.

\section*{Acknowledgments}
It is my pleasure to thank Ira Wasserman and Henry Tye for enjoyable
collaborations.
This work was supported in part by the Department of Energy.

\end{document}